# A methodology to measure the effectiveness of academic recruitment and turnover[1]

*Giovanni Abramo* (corresponding author)
  Laboratory for Studies of Research and Technology Transfer
  Institute for System Analysis and Computer Science (IASI-CNR)
  National Research Council of Italy
    ADDRESS: Via dei Taurini 19, 00185 Roma – ITALY
    tel. and fax +39 06 72597362, giovanni.abramo@uniroma2.it

*Ciriaco Andrea D'Angelo*
  Department of Engineering and Management
  University of Rome "Tor Vergata"
    ADDRESS: Via del Politecnico 1, 00133 Roma – ITALY
    tel. and fax +39 06 72597362, dangelo@dii.uniroma2.it

*Francesco Rosati*
  DTU Management Engineering
  Technical University of Denmark
    ADDRESS: Produktionstorvet Building 426
    2800 Kgs. Lyngby - Denmark
    tel +45 45256021, frro@dtu.dk

**Abstract**

We propose a method to measure the effectiveness of the recruitment and turnover of professors, in terms of their research performance. The method presented is applied to the case of Italian universities over the period 2008-2012. The work then analyses the correlation between the indicators of effectiveness used, and between the indicators and the universities' overall research performance. In countries that conduct regular national assessment exercises, the evaluation of effectiveness in recruitment and turnover could complement the overall research assessments. In particular, monitoring such parameters could assist in deterring favoritism, in countries exposed to such practices.

**Keywords**

*Research evaluation; recruitment; bibliometrics; universities; career advancement.*

# 1. Introduction

The quest for excellence in research and higher education systems is a high priority for national governments, faced with the current realities of the knowledge economy. Universities produce new knowledge and educate the future labor force, and thus represent a potential distinctive competence in strengthening the nation's competitive advantage. The European Union considers the contribution of higher education as central in giving the region "the most competitive economy and knowledge-based society of the 21st century". Nevertheless, the EU higher education system remains fragmented, and in many countries is hampered by a combination of excessive public control and scarce autonomy. In the recent years we have witnessed interventions by a growing number of governments to reinforce competitive market-like mechanisms and thus achieve greater effectiveness and efficiency (Jongbloed, 2004). In Northern Europe, the application of New Public Management in the academic sector, with emphasis on quasi-market competition, efficiency and performance audit practices, has lead to an overall increase in performance, together with greater differentiation among universities in the 1990s, and less so in the 2000s (Halffman and Leydesdorff, 2010). Expectations are also that, in the coming years, government funding will be distributed in an increasingly less uniform manner (Horta et al., 2008). In southern Europe, higher education systems are generally composed of public universities with relatively low autonomy, and are often characterized by weak overall performance with little differentiation among institutions (Van der Ploeg and Veugelers, 2008). In Italy, as a case in point, the variability of research productivity among universities (standardized citations per researcher in the same field) results as being much lower than that within individual institutions, which then result as rather homogeneous in terms of research performance (Abramo et al., 2012a).

A distinctive competence of prestigious universities is their ability in attracting and retaining the best professors. Other key competencies, such as success in attracting talented students and abundant resources, are a direct consequence of the quality of faculty. In these institutions, appointments to academic positions are normally managed through institutionally appointed *ad hoc* search committees, which advertise the competitions in international social networks and scientific journals. In contrast, a number of the European nations govern recruitment and advancement through the imposition of rigid procedures, regulated and in part enacted by a central bureaucracy. In nations with non-competitive higher education systems, and with generically high levels of corruption in public administration, this situation generates further exposure to phenomena of favoritism and nepotism in faculty recruitment and advancement. Italy is a case in point, as shown by empirical studies, judicial reports and media attention. Italian governments have intervened repeatedly to reduce the problem, with scarce success. In cases such as this, access to instruments that could measure the effectiveness of the universities' academic recruitment could serve as a deterrent against the activation and continuation of the practices of discrimination and favoritism.

In this work we present a methodology to measure, in comparative terms, the effectiveness of recruitment and turnover by universities. Since the single most important criterion in the selection of candidates to faculty positions is excellence in research, we simply compare the research performance of the new entrants to the average performance of their peers, in the same fields of research. Those universities that tend to recruit higher performing candidates will evidently be the most effective.

We conduct an analogous operation for the professors separating from faculty, in which we adopt the principle that the universities which succeed in disposing of low performers (i.e. retaining the higher performers) are those most effective in managing turnover. The current paper applies the methodology to the case of Italian universities, however the model should be of interest to all countries characterized by non-competitive higher education systems, as well as all those that adopt national research assessment exercises. The information gained through the methodology could be of interest in selective funding. It also contributes to greater symmetry of information between universities and stakeholders, and permits the universities and government to demonstrate to taxpayers that public money is effectively spent.

The next section of the paper describes the Italian context in terms of the higher education system and recruitment processes. Section 3 describes the methodology for measurement of scientific merit and presents the dataset for the analyses. In section 4 we analyze the Italian academic mobility in the period under observation. In section 5 we develop the indices of effectiveness for recruitment, turnover, and overall mobility, and present the resulting rankings from their application to Italian universities. The final section discusses the conclusions and implications.

## 2. The Italian higher education system

The Italian Ministry of Education, Universities and Research (MIUR) recognizes a total of 96 universities as having the authority to issue legally recognized degrees. Of these, 29 are small, private, special-focus universities, 67 are public and generally multi-disciplinary universities, scattered throughout Italy. 94.9% of faculty are employed in public universities. Public universities are largely financed by the government through mostly non-competitive allocation of funds. Until 2009 the core government funding was independent of merit, and distributed to universities in a manner intended to satisfy the needs of each and all equally, with respect to their size and research disciplines. It was only following the first national research evaluation exercise, conducted in the period 2004–6, that a minimal share, equivalent to 3.9% of total income, was assigned by the MIUR as a function of the assessment of research.

In keeping with the Humboldtian model, there are no 'teaching-only' universities in Italy, as all professors are required to carry out both research and teaching. National legislation includes a provision that each faculty member must devote a minimum of 350 hours per year to teaching activities.

Salaries are regulated at the central level and are calculated according to role (administrative, technical or professorial), rank within role (e.g. assistant, associate or full professor) and seniority. None of a professor's salary depends on merit. Moreover, as in all Italian public administration, dismissal of unproductive professors is unheard of. All new personnel enter the university system through public competitions (concorso), and career advancement depends on further public competitions. The entire legislative–administrative context creates a culture that is scarcely competitive, which is further associated with high levels of corruption and favoritism. According to The Global Competitiveness Report 2014-2015 (Schwab, 2014), Italy ranks 106[th] out of 144 countries in deterioration in the functioning of its institutions, and 134[th] in favoritism in decisions of government officials; while placing 54[th] out of 1 50 countries in the 2014

World Democracy Audit for corruption.[2] It is no surprise then that the nationally governed competitions for faculty positions have come under frequent fire, and that the Italian word "concorso" has gained international note for its implications of rigged competition, favoritism, nepotism and other unfair selection practices (Gerosa, 2001). Letters in prestigious journals such as The Lancet, Science and Nature (Garattini, 2001; Aiuti, 1994; Biggin, 1994; Amadori et al., 1992; Gaetani and Ferraris 1991; Fabbri, 1987), as well as entire monographs, (Perotti, 2008; Zagaria, 2007) continue to report on injustice in recruitment, including many cases that arrive in judicial proceedings. Abramo et al. (2014) investigated the 287 Italian associate professor competitions launched in 2008. The analysis showed several critical issues, particularly concerning unsuccessful candidates who outperformed the competition winners in terms of research productivity, as well as a number of competition winners who resulted as totally unproductive. Specifically, it emerged that 29% of the winners had productivity below the median of the performance distribution for their peers in the same field of research, and that 5.5% of the winners had not produced any significant advances in scientific knowledge. Almost half of individual competitions selected candidates who would go on to achieve below-median productivity in their field of research over the subsequent triennium. In a further work, the same authors attempted to discover the potential factors that could have contributed to the outcomes of the 2008 round of competitions (Abramo et al., 2015a). They identified that the main determinant of a candidate's success was not their scientific merit, but rather the number of years of service in the same university as the committee president. Where the candidate had actually cooperated in joint research work with the president, the probability of success again increased significantly. These results confirm what Zinovyeva and Bagues (2012) have demonstrated for the Spanish case, i.e. the existence of mechanisms of selection that are completely contrary to the intentions of the regulatory framework for the competitions.

   The overall result is a system of universities that are almost completely undifferentiated in research performance (Bonaccorsi and Cicero, 2015; Abramo et al., 2012a). Thus, universities are unable to attract significant numbers of talented foreign faculty: only 1.8% of research staff are foreign nationals. This is a system where every university has some share of top scientists, flanked by another share of absolute non-producers. Over the period 2004-2008, 6,640 (16.8%) of the 39,512 professors in the so called hard sciences, did not publish any scientific articles in the journals indexed by the Web of Science (WoS). Another 3,070 professors (7.8%) achieved publication, but their work was never cited (Abramo et al., 2013a). This means that 9,710 individuals (24.6%) had no impact on scientific progress measurable by bibliometric databases.[3] An almost equal 23% of professors alone produced 77% of the overall Italian scientific advancement. This 23% of 'top' faculty is not concentrated in a limited number of universities. Instead, it is dispersed more or less uniformly among all Italian universities, along with the unproductive academics, so that no single institution reaches the critical mass of excellence necessary to develop as a world-class university and to compete internationally (Abramo et al., 2012a). No wonder that there are no Italian universities among the top 150 of the Shanghai rankings or, with the only exception of the tiny School for Advanced Studies (Normale) in Pisa, in the top 200 of the "THE" rankings. Various Italian governments have made countless attempts to overcome the chronic social disease of favoritism, by changing the rules and procedures of the academic appointment system, however these have always failed. The adoption of regular assessment exercises of the research performance of new entrants in the academic system may act as a deterrent and help reduce practices of favoritism and nepotism.

---

[2] http://www.worldaudit.org/corruption.htm, last accessed on September 15, 2015.
[3] Researchers who we define as 'unproductive' may actually publish in journals not indexed by the WoS or codify the new knowledge produced in different forms, such as books, patents etc.

## 3. Method and data

In the Italian university system all professors are classified in one and only one field, named scientific disciplinary sector (SDS), 370 in all. SDSs are grouped into disciplines, named university disciplinary areas (UDAs), 14 in all.[4] Competitions for recruitment occur at SDS level.

In this work we will analyze all the cases of recruitment and turnover of professors in the so-called "bibliometric" disciplinary sectors (SDSs) of the Italian university system, over the period 2008-2012. The effectiveness of the universities in these processes will be measured by comparing the scientific performance of the new entrants and of the "leavers" with the performance of the incumbents and of the entire Italian academic staff. Scientific performance is measured by means of a bibliometric indicators based on the publications indexed in the WoS. The limitation of the observations to the "bibliometrics fields" ensures the representativity of publications as a proxy of research output. We define an SDS as bibliometric if at least 50% of professors belonging to that SDS produced at least one publication over 2008-2012 period. Given the difficulty of appropriate application of bibliometric techniques, the professors of other SDSs (mainly in the Arts and Humanities and the Social sciences) are excluded from the analyses.

Scientific performance is measured through an indicator of productivity. Most bibliometricians define productivity as the number of publications in the period under observation. Because publications have different values (impact), we prefer to adopt a more meaningful definition of productivity: the value of output per unit value of labor, all other production factors being equal. The latter recognizes that the publications embedding new knowledge have a different value or impact on scientific advancement, which can be approximated with citations or journal impact factors. Provided that there is an adequate citation window (at least two years) the use of citations is always preferable (Abramo et al., 2010). Because citation behavior varies by field, we standardize the citations for each publication with respect to the average of the distribution of citations for all the cited Italian publications indexed in the same year and the same WoS subject category.[5] Furthermore, research projects frequently involve a team of professors, which is registered in the co-authorship of publications. In this case we account for the fractional contributions of scientists to outputs, which is sometimes further signaled by the position of the authors in the list of authors. Unfortunately, relevant data on the production factors available to each professor are not known in Italy. We assume then that the same resources are available to all professors within the same field, and that the hours devoted to educational activities are more or less the same for all professors.

---

[4] The complete list is accessible on http://attiministeriali.miur.it/UserFiles/115.htm, last accessed on September 15, 2015.
[5] Abramo et al. (2012b) demonstrated that the average of the distribution of citations received for all cited publications of the same year and subject category is the best-performing scaling factor.

At the individual level, we can measure a proxy of the average yearly productivity, termed fractional scientific strength (*FSS*), as follows:[6]

$$FSS = \frac{1}{t}\sum_{i=1}^{N}\frac{c_i}{\bar{c}}f_i$$

[1]

Where:

$t$ = number of years of work by researcher in period under observation
$N$ = number of publications by researcher in period under observation
$c_i$ = citations received by publication, $i$
$\bar{c}$ = average of distribution of citations received for all cited publications in same year and subject category of publication, $i$
$f_i$ = fractional contribution of researcher to publication, $i$.

As for fractional contribution, we use a flexible co-authorship credit allocation scheme (Kim and Diesner, 2014). In Italy, in most fields the practice is to place the authors in simple alphabetical order: in this case the fractional contribution simply equals the inverse of the number of authors. For the life sciences, widespread and consolidated practice in Italy is for the authors to indicate the various contributions to the published research by the order of the names in the listing of the authors. For this reason in life science SDSs, as required by the relevant scientific community, we use a network based flexible allocation scheme, giving different weights to each co-author according to their position in the list of authors and the character of the co-authorship (intra-mural or extra-mural) (Abramo et al., 2013b). If the first and last authors belong to the same university, 40% of the citation is attributed to each of them, the remaining 20% is divided among all other authors. If the first two and last two authors belong to different universities, 30% of the citation is attributed to the first and last authors, 15% of the citation is attributed to the second and last authors but one, the remaining 10% is divided among all others.[7]

Data on the faculty at each university and their SDS classification were extracted from the database on Italian university personnel, maintained by the MIUR[8]. The bibliometric dataset used to measure *FSS* is extracted from the Italian Observatory of Public Research, a database developed and maintained by the present authors and derived under license from the Thomson Reuters WoS. Beginning from the raw data of the WoS, and applying a complex algorithm to reconcile the author's affiliation and disambiguation of the true identity of the authors, each publication (article, article review and conference proceeding) is attributed to the university scientist or scientists that produced it (D'Angelo et al., 2011). Thanks to this algorithm we can produce rankings of research productivity at the individual level, on a national scale. Based on the value of *FSS* we obtain, for each SDS and academic rank, a ranking list expressed either on a percentile scale of 0–100 (worst to best) or as the *FSS* ratio to the average, to compare with the performance of all Italian academics of the same SDS and academic rank. We conduct the comparisons within the same academic rank, since the average professorial salary increases with academic rank, meaning that the cost of an assistant professors is less than that of a full professor. In addition, research productivity generally increases with academic rank (Abramo et al., 2011). In the cases where we wish to compare the performance of the professors of different academic rank, we can do so through normalizing by the average salary of the ranks.

---

[6] A more extensive theoretical dissertation on how to operationalize the measurement of productivity can be found in Abramo and D'Angelo (2014).
[7] The weightings were assigned following advice from senior Italian professors in the life sciences. The values could be changed to suit different practices in other national contexts.
[8] http://cercauniversita.cineca.it/php5/docenti/cerca.php, last accessed on September 15, 2015.

In measuring the effectiveness of recruitment we can adopt two perspectives. The first is to inquire into the capacity of the university to attract the best professors possible on the basis of the merit of the existing faculty, meaning relative to the scientific accomplishment or "reputation" of the university. We call this perspective "internal", and for this we compare the average performance of the incoming professors with that of the SDS faculty they are hired into. The second perspective, which we call "external", ignores the scientific reputation of the incumbent faculty, and so compares the average performance of the recruited professors with that of all Italian professors in the same SDSs and academic rank. Each perspective is analyzed using two indicators, for a total of four indicators of effectiveness of recruitment.

For the internal perspective we use the two indicators:

R 1.1: For each recruit we calculate the ratio of their FSS to the average FSS of the faculty in the given university and SDS, $\overline{FSS^I}$, (internal productivity)[9,10]. The average of these ratios calculated on the total of new recruits (N), gives us the measure of the indicator R 1.1 for each university. In formulae:

$$R\ 1.1 = \frac{1}{N}\sum_{j=1}^{N} \frac{FSS_j}{\overline{FSS^I}}$$

[2]

This indicator expresses the universities capability to attract talented professors, with respect to its own potential. For the more productive universities this ratio will tend to be somewhat lesser: a university with a faculty of only top professors can at maximum achieve a ratio of 1. Less productive universities can instead have very high ratios. A ratio greater than 1 indicates that the new entrants have increased the university's average performance.

R 1.2: This indicator is given by the percentage of recruits to a university who have FSS higher than the average of the FSSs for the faculty of the same SDS and academic rank, on the total of recruits. In formulae:

$$R\ 1.2 = \frac{1}{N}\sum_{j=1}^{N} x_j$$

[3]

with $x_j = 1$ if $FSS_j > \overline{FSS^I}$; otherwise 0.

This second indicator compensates for potential effects from outliers, which can affect indicator R 1.1.

---

[9] Since the number of incumbent professors in the same SDS and academic rank could be very low (or nil), at a given university, we compare to the performance of all the ranks, after normalizing the FSS by average salary per rank (Abramo and D'Angelo, 2014).

[10] In theory, it is possible that all incumbents are unproductive ($\overline{FSS^I} = 0$). To circumvent the mathematical problem (R 1.1 = ∞), one could assign to $\overline{FSS^I}$ the lowest value of the productivity distribution, and increase the numerator by the same value. As a matter of fact, this case never occurred in our analysis.

For the measure of the effectiveness of recruitment from the external perspective, we use two indicators:

R 2.1: For each recruit we first calculate the ratio of her or his FSS to the average FSS of all the professors of equal academic rank in the same SDS within the entire Italian university system ($\overline{FSS^E}$, external productivity).[11] Indicator R 2.1 for each university is given by the average of these ratios calculated on the total of new recruits (N). In formulae:

$$R\ 2.1 = \frac{1}{N} \sum_{j=1}^{N} \frac{FSS_j}{\overline{FSS^E}}$$

[4]

R 2.2: This indicator is given by the percentage of recruits to a university, with FSS greater than the average of the FSSs of the same rank and SDS in the entire Italian university system, on the total of all entrants (N). In formulae:

$$R\ 2.2 = \frac{1}{N} \sum_{j=1}^{N} y_j$$

[5]

with $y_j = 1$ if $FSSE_j > \overline{FSS^E}$; otherwise 0.

For the measure of the effectiveness of turnover, meaning the university's ability to retain the best professors, we again adopt two perspectives and four indicators. From the internal perspective, we compare the performance of the professors separated from the university with the incumbents currently at the same institution. From the external perspective, we compare the performance of the separated professors with the incumbents of the entire nation.

For the internal perspective, we use the two indicators:

T 1.1: For each separated professor we first calculate the ratio between the average FSS of the faculty in the same SDS, expressed $\overline{FSS^I}$ (internal productivity), and the leaver's FSS. The average of these ratios calculated on the total of "leavers" (P), gives the measure of the indicator T 1.1 for every university. In formulae:

$$T\ 1.1 = \frac{1}{P} \sum_{j=1}^{P} \frac{\overline{FSS^I}}{FSS_j}$$

[6]

Indicator T 1.1[12] expresses the extent to which a university is capable of retaining the best professors, relative to its own potential.

T 1.2: This indicator is given by the percentage of leavers from a university with FSS less than the $\overline{FSS^I}$ of faculty in the same SDS, in the total of the university's leavers

---

[11] In this case, having a greater number of observations, it is possible to compare FSS between the professors of the individual academic ranks; for this there is no need to normalize by average salary of rank.

[12] In theory, it is possible that the leaver is unproductive ($FSS_j = 0$). To circumvent the mathematical problem (T 1.1 = ∞), one could assign to $FSS_j$ the lowest value of the productivity distribution, and increase the numerator by the same value. Although possible in theory, it is difficult to imagine that an assistant or associate professor in University A is hired by University B at a higher rank, with 0 productivity in the previous five years (our population of leavers does not include retired professors or professors exiting the academic system, see next). As a matter of fact, this case never occurred in our analysis.

(P). In formulae:

$$\text{T 1.2} = \frac{1}{P}\sum_{j=1}^{P} x_j$$

[7]

with $x_j = 1$ if $FSS_j < \overline{FSS^I}$; otherwise 0.

For the external perspective, we use the following indicators:

T 2.1: For each separated professor we calculate the ratio of the average FSS of all the professors of the Italian university system of the same academic rank and SDS, $\overline{FSS^E}$ (external productivity), to the leaver's FSS[13]. The average of these ratios calculated on the total of the university's leavers (P), gives us the measure of the indicator T 2.1 for each institution. In formulae:

$$\text{T 2.1} = \frac{1}{P}\sum_{j=1}^{P} \frac{\overline{FSS^E}}{FSS_j}$$

[8]

T 2.2: This indicator is given by the percentage of leavers from a university, with FSS less than the average FSS of all the professors of the Italian university system of the same SDS and academic rank, on the total of leavers (P). In formulae:

$$\text{T 2.2} = \frac{1}{P}\sum_{j=1}^{P} y_j$$

[9]

with $y_j = 1$ if $FSS_j < \overline{FSS^E}$; otherwise 0.

Finally, it is possible to calculate the overall effectiveness of the recruitment and turnover of a university (mobility), as the weighted average of the two. In formulae, for each indicator:

$$\text{M 1.1} = \frac{N \cdot R\,1.1 + P \cdot T\,1.1}{N + P}$$

[10]

$$\text{M 1.2} = \frac{N \cdot R\,1.2 + P \cdot T\,1.2}{N + P}$$

[11]

$$\text{M 2.1} = \frac{N \cdot R\,2.1 + P \cdot T\,2.1}{N + P}$$

[12]

$$\text{M 2.2} = \frac{N \cdot R\,2.2 + P \cdot T\,2.2}{N + P}$$

[13]

The usual warrants in the application of the bibliometric methodology and interpretation of results apply. Limits and assumptions are embedded in both the FSS indicator and the methodology to assess the ability of recruitment and turnover. As for the factors that introduce uncertainty (and bias) in bibliometric indicators, and how to measure (reduce) them, we refer the reader to Abramo et al. (2015b). The main limitation of the methodology is the assumption that the research performance of professors explains their overall academic merit.

---
[13] See footnote 13.

## 4. Mobility in Italian universities, 2008-2012

Italian labor law is exceptionally protective of public-sector workers. In the collective imagination, a position in the public sector means permanent, undeniable employment. Indeed, dismissal over performance issues is practically impossible. Given this tradition, and the further normative framework for the university sector, academic turnover in Italy is primarily voluntary, in general taking place for the personal motives of professors rather than for any pressure that the university might exert. Given the administrative context of Italian universities, measuring the effectiveness of turnover is thus more demonstrative of an inherited situation, rather than functioning in determining the merit/demerit of the current management. Currently, the vast majority of professors that leave a position do so to enter retirement, while very few exit the academic system for good. Few others transfer to another university to a higher rank. Transfers are still subject to the same public competitions as other new hires. In the period under observation, regulations for the competitions required the appointment of committees to judge the candidates' merits. Each committee was to be composed of five full professors belonging to the SDS for which the position was open. One member, the president, was designated by the university holding the competition and the other four were drawn at random from a short list of other full professors in the SDS concerned. The short list was in turn established by national voting among all the full professors of the SDS. After individual and joint judgments on all the candidates, the committee was required to vote to select two winners.[14]

From the database on Italian university personnel (Section 3), we extract all professors recruited in the period 2008-2012. To ensure robustness in the assessment of their research productivity, we exclude those with less than three years on faculty (Abramo et al., 2012c). The total number of recruits who entered faculty for at least three years in the observed period was 2,259, hired by 88 out of 96 universities. Of these professors, 1,781 (75%) were new entrants to the academic system, distributed by academic rank as 1,724 (96.8%) assistant professors, 43 (2.4%) associate professors, and 14 full professors (0.8%). The distribution confirms that Italian universities are a closed shop, meaning that it is highly exceptional to recruit any researcher to full or even associate professor from any context outside the assistant professors of the national system. The 1,781 new hires are equivalent to 5.8% of the incumbent professors (30,913). The remaining 478 professors have transferred from one university to another, meaning that they are simultaneously cases of unforced turnover for their departed university and of recruitment for the destination university. The analyses of turnovers will be carried out only on this subpopulation, although there are other cases of professors leaving the academic system. However, among these we are unable to distinguish those who retired from those who left for other motives, meaning that we cannot consider the latter for purposes of measuring the effectiveness of turnover.

The base assumptions for the dataset mean that the number of observations of new hires (2,259; 7.3% of incumbents) is five times the number of turnovers (478; 1.5% of incumbents). In related manner, while there are 70 universities with at least four recruits in the five-year period, there are only 49 with this many turnovers. The incidence of recruits in incumbents reaches a maximum of 28.4%, for the University of Napoli 'Parthenope'; the maximum incidence of turnovers (16.7%) is registered at University of Molise. Table 1 shows the data for incumbents, recruits, turnover and total mobility, by UDA (bibliometric SDSs only). Concerning new hires, the most active UDA is Medicine, with 536 entries (6.2% of incumbents). The same UDA also shows the greatest number of separations (123; equal to 1.4% of incumbents) and the greatest total mobility (659; 7.6% of incumbents). This lead is clearly linked to the size of the UDA, which outclasses all the others for number of incumbents (8,641). However in relative terms, the most dynamic UDA is Economics and statistics, with a number of recruits (231) equal to 17.0% of incumbents, and of separations (88) equal to 6.5% of incumbents, for a total mobility concerning 319 professors, or 23.5% of the UDA research staff.

---

[14] The committees could also indicate a single winner or reject all the applicants, however such events were very rare.

| UDA | Incumbents | Recruits | Turnover | Total Mobility |
|---|---|---|---|---|
| Mathematics and computer science | 2,648 | 224 (8,5) | 83 (3,1) | 307 (11,6) |
| Physics | 1,894 | 111 (5,9) | 13 (0,7) | 124 (6,5) |
| Chemistry | 2,540 | 135 (5,3) | 13 (0,5) | 148 (5,8) |
| Earth sciences | 905 | 64 (7,1) | 15 (1,7) | 79 (8,7) |
| Biology | 4,157 | 278 (6,7) | 36 (0,9) | 314 (7,6) |
| Medicine | 8,641 | 536 (6,2) | 123 (1,4) | 659 (7,6) |
| Agricultural and veterinary sciences | 2,392 | 132 (5,5) | 16 (0,7) | 148 (6,2) |
| Civil engineering and architecture | 1,283 | 105 (8,2) | 17 (1,3) | 122 (9,5) |
| Industrial and information engineering | 4,341 | 332 (7,6) | 49 (1,1) | 381 (8,8) |
| Pedagogy and psychology | 756 | 111 (14,7) | 25 (3,3) | 136 (18,0) |
| Economics and statistics | 1,356 | 231 (17,0) | 88 (6,5) | 319 (23,5) |
| Total | 30,913 | 2,259 (7,3) | 478 (1,5) | 2,737 (8,9) |

*Table 1: Number of incumbents, recruits, separations and total mobility (in brackets % of incumbents), by UDA*

## 5. Results

In this section we evaluate the effectiveness of recruitment, turnover and overall mobility of the universities, based on the average performance of the recruited and separated professors as compared to that of both the incumbents of their same university and of the entire population of Italian professors. Finally we analyze the correlation between the different indicators of effectiveness, and between the indicators and the research productivity of universities.

Tables 2, 3 and 4 present the indicators of effectiveness of recruitment, departures and overall mobility for all the universities of the dataset, omitting those with less than a total of four new hires, four leaving faculty and four incumbents in the bibliometric SDSs. Given these exclusions, the number of universities analyzed drops from 88 to 49.

The analyses of effectiveness of recruitment show that, in terms of "interior" indicator R 1.1, the most effective institution is the University of Parma, and the least effective is the University of Molise. This university is also last in the ranking list for the other three indicators. However in terms of indicator R 1.2, the best performing institution is instead the University of Salento. By the indicator R 2.1, the best performing university is the University of Teramo, while by indicator R 2.2 it is the University of Pavia that tops the ranking.

The analyses for effectiveness of turnover shows that regardless of the indicator, the University of Catania always ranks at the top. For indicator T 1.2, this university is in fact one of three sharing top place, together with those of Basilicata and Trento, while for T 2.2, it is only the universities of Catania and Basilicata in top spot. At the opposite extreme, the University of Sassari is last in the rankings for both of the external indicators, T 2.1 and T 2.2, and also third last for the other two indicators. The Trieste SISSA and the University of Salento share last place in the rankings for both T 2.1 and T 2.2.

The analysis of the overall effectiveness by indicator M 1.1 indicates the first ranked is the University of Parma, followed by Perugia and Sassari, however, contrastingly, Sassari places last in the rankings for the other three indicators. Indeed, overall effectiveness seems to be the dimension of analysis with the lowest correlation among the set of four different indicators considered. For example Messina is in $86^{th}$ percentile for M 1.1 but leaps to eighth for M 1.2 and fourth for M 2.1. The university of Salento is in the fourth percentile for M 2.2, but $98^{th}$ for M 1.2. An exception to this instability is Trieste SISSA, which uniformly places last in the rankings, independent of the indicator. Given these fluctuations, we will now attempt to probe deeper, analyzing the correlation between each indicator, as well as between each of these indicators and the universities' productivity. Table 5 shows the results of the correlations, where productivity is calculated over the period 2008-2012, through the use of the individual professorial FSS ratios[15] (Abramo and D'Angelo, 2014).

From the table, we observe that the indicators of a single dimension of effectiveness are generally correlated within their groups, and that the strongest correlation is seen for the four indicators of effectiveness of turnover. In contrast, there does not appear to be correlation between the indicators of effectiveness of recruitment and those for turnover.

However, there are very strong correlations between the indicators of effectiveness of recruitment and those for overall effectiveness (Spearman ρ for correlation between R 1.1 and M 1.1 is 0.92; that between R 1.2 and M 1.2 is 0.82), and in lesser measure, between the indicators of effectiveness of turnover and those for overall effectiveness (Spearman ρ 0.58 between T 2.1 and M 2.1 and 0.49 between T 2.2 and M 2.2). This result is clearly influenced by the fact that the study dataset is composed of a greater number of new hires than separated professors. From the analyses, it further emerges that there is a lack of correlation between the productivity of the universities and their ability to retain talented scholars, as well as a lack of correlation between productivity and the overall effectiveness of mobility. However, we do note a substantial and significant correlation between the universities' productivity and the indicators of effectiveness in recruitment, from the external perspective (R 2.1 and R 2.2, Spearman ρ respectively 0.47 and 0.45). This is explained by the fact that the top performers tend to orient their employment seeking towards the more productive universities, and/or that these universities concentrate their efforts on recruiting the best candidates.

---

[15] The FSS ratio is the ratio of the individual FSS to the average FSS of all professors of the same SDS and academic rank.

| University | R 1.1 | rank % | R 1.2 | rank % | R 2.1 | rank % | R 2.2 | rank % |
|---|---|---|---|---|---|---|---|---|
| Parma | 13.27 | 100 | 50.0 | 54 | 1.17 | 36 | 38.5 | 44 |
| Sassari | 11.49 | 98 | 44.4 | 21 | 0.51 | 2 | 16.7 | 2 |
| Perugia | 10.06 | 96 | 50.8 | 56 | 1.36 | 54 | 40.3 | 52 |
| Milan Bicocca | 8.28 | 94 | 51.9 | 58 | 1.41 | 63 | 44.4 | 58 |
| Brescia | 8.10 | 92 | 47.8 | 36 | 1.13 | 31 | 48.3 | 69 |
| Basilicata | 4.94 | 90 | 52.9 | 63 | 1.06 | 23 | 30.4 | 17 |
| Messina | 4.17 | 88 | 37.3 | 8 | 0.86 | 11 | 24.1 | 6 |
| Siena | 3.70 | 86 | 68.2 | 98 | 1.28 | 44 | 52.0 | 79 |
| Verona | 3.63 | 83 | 52.9 | 63 | 1.49 | 69 | 52.8 | 83 |
| Piemonte Orientale A. Avogadro | 3.53 | 81 | 53.3 | 65 | 2.60 | 98 | 52.9 | 86 |
| Salerno | 3.32 | 79 | 45.2 | 23 | 0.95 | 15 | 31.5 | 21 |
| L'Aquila | 3.22 | 77 | 53.8 | 71 | 1.79 | 88 | 39.3 | 48 |
| Pavia | 3.03 | 75 | 65.1 | 94 | 2.19 | 92 | 72.7 | 100 |
| Trento | 2.92 | 73 | 44.4 | 21 | 1.79 | 90 | 47.8 | 65 |
| Cassino | 2.80 | 71 | 50.0 | 54 | 1.56 | 75 | 35.7 | 38 |
| Mediterranea di Reggio Calabria | 2.72 | 69 | 25.0 | 4 | 0.61 | 4 | 20.0 | 4 |
| Pisa | 2.55 | 67 | 63.8 | 92 | 1.60 | 77 | 46.9 | 63 |
| Catania | 2.46 | 65 | 57.5 | 83 | 1.04 | 17 | 33.3 | 33 |
| Magna Grecia di Catanzaro | 2.41 | 63 | 46.2 | 31 | 1.28 | 48 | 56.7 | 96 |
| Naples Second Napoli | 2.34 | 61 | 49.0 | 44 | 1.68 | 83 | 44.2 | 54 |
| Rome 'La Sapienza' | 2.31 | 58 | 48.8 | 40 | 1.39 | 61 | 40.0 | 50 |
| Salento | 2.15 | 56 | 100.0 | 100 | 1.68 | 86 | 50.0 | 75 |
| Modena and Reggio Emilia | 2.10 | 54 | 59.4 | 88 | 1.05 | 19 | 36.1 | 40 |
| Milan | 2.04 | 52 | 48.9 | 42 | 2.25 | 94 | 53.3 | 90 |
| Gabriele D'Annunzio | 2.00 | 50 | 57.1 | 81 | 1.37 | 56 | 44.4 | 58 |
| Florence | 1.92 | 48 | 54.1 | 73 | 1.52 | 73 | 45.9 | 61 |
| Genoa | 1.92 | 46 | 53.6 | 67 | 0.90 | 13 | 27.3 | 13 |
| Ferrara | 1.85 | 44 | 50.0 | 54 | 1.26 | 42 | 33.3 | 33 |
| Padua | 1.79 | 42 | 55.4 | 75 | 2.27 | 96 | 60.2 | 98 |
| Urbino 'Carlo Bo' | 1.78 | 40 | 41.7 | 15 | 1.19 | 38 | 33.3 | 33 |
| Calabria | 1.74 | 38 | 59.3 | 86 | 1.08 | 27 | 37.9 | 42 |
| Milan Polytechnic | 1.74 | 36 | 56.2 | 79 | 1.48 | 67 | 53.3 | 90 |
| Cagliari | 1.72 | 33 | 66.7 | 96 | 1.43 | 65 | 55.6 | 94 |
| Turin Polytechnic | 1.68 | 31 | 62.9 | 90 | 1.28 | 46 | 52.8 | 83 |
| Trieste | 1.65 | 29 | 53.8 | 71 | 1.52 | 71 | 35.7 | 38 |
| Teramo | 1.63 | 27 | 50.0 | 54 | 2.76 | 100 | 54.5 | 92 |
| Udine | 1.61 | 25 | 45.5 | 25 | 1.16 | 33 | 29.7 | 15 |
| Napoli 'Federico II' | 1.58 | 23 | 49.1 | 46 | 1.09 | 29 | 38.6 | 46 |
| Foggia | 1.50 | 21 | 45.7 | 27 | 1.23 | 40 | 47.9 | 67 |
| Bari | 1.47 | 19 | 47.5 | 33 | 1.06 | 21 | 31.7 | 23 |
| Rome 'Tor Vergata' | 1.45 | 17 | 40.0 | 13 | 1.07 | 25 | 33.3 | 33 |
| Turin | 1.38 | 15 | 48.6 | 38 | 1.39 | 58 | 49.3 | 71 |
| Palermo | 1.37 | 13 | 38.6 | 11 | 0.83 | 8 | 25.0 | 11 |
| Cattolica del Sacro Cuore | 1.29 | 11 | 46.2 | 31 | 0.66 | 6 | 31.3 | 19 |
| Bologna | 1.19 | 8 | 42.7 | 17 | 1.29 | 50 | 51.3 | 77 |
| Trieste SISSA | 1.16 | 6 | 25.0 | 4 | 1.35 | 52 | 25.0 | 11 |
| 'Campus Bio-medico' | 1.15 | 4 | 55.6 | 77 | 1.63 | 79 | 50.0 | 75 |
| Insubria | 0.91 | 2 | 27.8 | 6 | 1.67 | 81 | 33.3 | 33 |
| Molise | 0.36 | 0 | 20.0 | 0 | 0.42 | 0 | 0.0 | 0 |
| Total | 2.79 | - | 50.1 | - | 1.37 | - | 42.2 | - |

*Table 2: Effectiveness of recruitment by Italian universities (sorted by R 1.1)*

| University | T 1.1 | rank % | T 1.2 | rank % | T 2.1 | rank % | T 2.2 | rank % |
|---|---|---|---|---|---|---|---|---|
| Catania | 42.93 | 100 | 100.0 | 100 | 33.12 | 100 | 100.0 | 100 |
| Pisa | 14.17 | 98 | 33.3 | 36 | 2.94 | 58 | 40.0 | 17 |
| Palermo | 10.18 | 96 | 33.3 | 36 | 3.01 | 61 | 40.0 | 17 |
| della Basilicata | 9.31 | 94 | 100.0 | 100 | 9.18 | 96 | 100.0 | 100 |
| Milan Bicocca | 5.77 | 92 | 42.9 | 52 | 3.25 | 69 | 50.0 | 36 |
| Florence | 4.98 | 90 | 42.9 | 52 | 4.56 | 86 | 33.3 | 11 |
| Trento | 3.86 | 88 | 100.0 | 100 | 2.88 | 56 | 75.0 | 77 |
| Udine | 3.00 | 86 | 62.5 | 79 | 2.56 | 50 | 58.3 | 48 |
| Catanzaro 'Magna Grecia' | 2.96 | 83 | 33.3 | 36 | 4.57 | 88 | 71.4 | 67 |
| Ferrara | 2.95 | 81 | 66.7 | 90 | 4.64 | 90 | 70.0 | 63 |
| Insubria | 2.75 | 79 | 60.0 | 75 | 7.08 | 94 | 75.0 | 77 |
| Trieste | 2.67 | 77 | 60.0 | 75 | 2.43 | 48 | 80.0 | 88 |
| Rome 'La Sapienza' | 2.37 | 75 | 40.0 | 46 | 3.41 | 73 | 66.7 | 58 |
| Verona | 2.35 | 73 | 23.1 | 13 | 2.00 | 44 | 35.7 | 13 |
| Brescia | 2.15 | 71 | 75.0 | 94 | 1.20 | 19 | 80.0 | 88 |
| Turin Polytechnic | 2.07 | 69 | 75.0 | 94 | 1.70 | 36 | 55.6 | 42 |
| Molise | 1.96 | 67 | 53.3 | 69 | 4.99 | 92 | 75.0 | 77 |
| Siena | 1.90 | 65 | 66.7 | 90 | 2.81 | 54 | 70.0 | 63 |
| 'Campus Bio-medico' | 1.85 | 63 | 50.0 | 67 | 3.12 | 65 | 75.0 | 77 |
| Cattolica del Sacro Cuore | 1.84 | 61 | 60.0 | 75 | 2.57 | 52 | 75.0 | 77 |
| Gabriele D'Annunzio | 1.77 | 58 | 50.0 | 67 | 1.70 | 38 | 60.0 | 56 |
| Mediterranea di Reggio Calabria | 1.77 | 56 | 50.0 | 67 | 2.01 | 46 | 80.0 | 88 |
| Salerno | 1.75 | 54 | 50.0 | 67 | 3.55 | 75 | 100.0 | 100 |
| Bologna | 1.67 | 52 | 42.9 | 52 | 4.12 | 81 | 58.8 | 50 |
| Genoa | 1.58 | 50 | 37.5 | 40 | 3.70 | 79 | 55.6 | 42 |
| Parma | 1.36 | 48 | 50.0 | 67 | 1.22 | 21 | 44.4 | 25 |
| Naples 'Second' | 1.31 | 46 | 66.7 | 90 | 15.92 | 98 | 80.0 | 88 |
| Cassino | 1.27 | 44 | 62.5 | 79 | 1.49 | 29 | 83.3 | 92 |
| Calabria | 1.25 | 42 | 20.0 | 11 | 1.79 | 40 | 100.0 | 100 |
| Milan | 1.24 | 40 | 28.6 | 21 | 3.25 | 67 | 44.4 | 25 |
| Bari | 1.23 | 38 | 50.0 | 67 | 1.87 | 42 | 43.8 | 21 |
| Perugia | 1.14 | 36 | 66.7 | 90 | 1.46 | 27 | 71.4 | 67 |
| Naples 'Federico II' | 1.11 | 33 | 27.3 | 19 | 3.65 | 77 | 57.1 | 46 |
| Piemonte Orientale A. Avogadro | 1.02 | 31 | 27.3 | 19 | 1.19 | 15 | 45.5 | 27 |
| Teramo | 1.01 | 29 | 63.6 | 81 | 1.59 | 31 | 83.3 | 92 |
| Torino | 0.99 | 27 | 38.5 | 42 | 1.67 | 33 | 53.3 | 38 |
| Foggia | 0.99 | 25 | 40.0 | 46 | 4.48 | 83 | 57.1 | 46 |
| Padua | 0.94 | 23 | 33.3 | 36 | 1.28 | 25 | 60.0 | 56 |
| Messina | 0.87 | 21 | 33.3 | 36 | 1.25 | 23 | 50.0 | 36 |
| Modena e Reggio Emilia | 0.77 | 19 | 33.3 | 36 | 0.92 | 6 | 28.6 | 8 |
| del Salento | 0.75 | 17 | 33.3 | 36 | 0.38 | 0 | 0.0 | 2 |
| L'Aquila | 0.71 | 15 | 25.0 | 15 | 1.13 | 13 | 50.0 | 36 |
| Milan Polytechnic | 0.63 | 13 | 50.0 | 67 | 0.94 | 8 | 50.0 | 36 |
| Trieste SISSA | 0.61 | 11 | 20.0 | 11 | 0.50 | 2 | 0.0 | 2 |
| Urbino 'Carlo Bo' | 0.59 | 8 | 0.0 | 4 | 1.00 | 11 | 25.0 | 6 |
| Pavia | 0.43 | 6 | 0.0 | 4 | 1.19 | 17 | 60.0 | 56 |
| Cagliari | 0.38 | 4 | 37.5 | 40 | 3.01 | 63 | 80.0 | 88 |
| Rome 'Tor Vergata' | 0.27 | 2 | 11.1 | 6 | 3.26 | 71 | 41.7 | 19 |
| Sassari | 0.10 | 0 | 0.0 | 4 | 0.58 | 4 | 20.0 | 4 |
| Total | 2.33 | - | 45.9 | - | 3.45 | - | 59.8 | - |

*Table 3: Effectiveness of turnover by Italian universities (sorted by T 1.1)*

| University | M 1.1 | rank % | M 1.2 | rank % | M 2.1 | rank % | M 2.2 | rank % |
|---|---|---|---|---|---|---|---|---|
| Parma | 11.44 | 100 | 50.0 | 52 | 1.19 | 21 | 40.0 | 27 |
| Perugia | 9.42 | 98 | 52.1 | 58 | 1.37 | 33 | 43.2 | 36 |
| Sassari | 9.22 | 96 | 33.3 | 4 | 0.53 | 0 | 17.6 | 2 |
| Milano Bicocca | 8.05 | 94 | 50.8 | 54 | 1.63 | 50 | 45.2 | 44 |
| Brescia | 7.35 | 92 | 51.9 | 56 | 1.14 | 17 | 52.9 | 77 |
| Basilicata | 5.77 | 90 | 63.6 | 94 | 2.51 | 92 | 44.8 | 42 |
| Catania | 4.39 | 88 | 60.5 | 88 | 3.23 | 96 | 39.1 | 25 |
| Messina | 3.98 | 86 | 37.0 | 8 | 0.89 | 4 | 25.8 | 8 |
| Verona | 3.26 | 83 | 44.7 | 23 | 1.63 | 52 | 48.0 | 58 |
| Pisa | 3.24 | 81 | 62.0 | 92 | 1.72 | 58 | 46.3 | 50 |
| Salerno | 3.14 | 79 | 45.7 | 31 | 1.10 | 13 | 35.1 | 19 |
| Trento | 3.09 | 77 | 54.5 | 73 | 1.97 | 71 | 51.9 | 71 |
| Siena | 2.89 | 75 | 67.5 | 100 | 1.94 | 69 | 60.0 | 92 |
| L'Aquila | 2.89 | 73 | 50.0 | 52 | 1.68 | 56 | 41.2 | 31 |
| Pavia | 2.86 | 71 | 60.9 | 90 | 2.09 | 79 | 71.4 | 100 |
| Catanzaro 'Magna Grecia' | 2.52 | 69 | 43.8 | 21 | 1.91 | 65 | 59.5 | 90 |
| Piemonte Orientale A. Avogadro | 2.43 | 67 | 42.3 | 17 | 2.05 | 77 | 50.0 | 67 |
| Florence | 2.41 | 65 | 52.3 | 61 | 2.13 | 83 | 43.5 | 38 |
| Cassino | 2.36 | 63 | 55.6 | 81 | 1.53 | 48 | 57.7 | 88 |
| Rome 'La Sapienza' | 2.32 | 61 | 48.6 | 40 | 1.46 | 40 | 40.9 | 29 |
| Naples 'Second' | 2.28 | 58 | 50.0 | 52 | 4.53 | 100 | 52.2 | 73 |
| Mediterranea di Reggio Calabria | 2.25 | 56 | 37.5 | 13 | 1.31 | 27 | 50.0 | 67 |
| Ferrara | 2.09 | 54 | 54.1 | 71 | 2.01 | 73 | 41.3 | 33 |
| Gabriele D'Annunzio | 1.96 | 52 | 55.8 | 83 | 1.43 | 36 | 47.8 | 56 |
| Milan | 1.94 | 50 | 46.2 | 33 | 2.42 | 88 | 51.9 | 71 |
| Trieste | 1.93 | 48 | 55.6 | 81 | 1.76 | 61 | 47.4 | 54 |
| Modena and Reggio Emilia | 1.89 | 46 | 55.3 | 77 | 1.03 | 8 | 34.9 | 17 |
| Udine | 1.89 | 44 | 48.8 | 42 | 1.53 | 46 | 36.7 | 23 |
| Genoa | 1.84 | 42 | 50.0 | 52 | 1.51 | 44 | 33.3 | 13 |
| Turin Polytechnic | 1.75 | 40 | 65.1 | 96 | 1.36 | 31 | 53.3 | 81 |
| Milan Polytechnic | 1.73 | 38 | 56.1 | 86 | 1.47 | 42 | 53.2 | 79 |
| Padua | 1.71 | 36 | 53.3 | 67 | 2.16 | 86 | 60.2 | 94 |
| Calabria | 1.68 | 33 | 53.1 | 65 | 1.18 | 19 | 47.1 | 52 |
| Palermo | 1.66 | 31 | 38.5 | 15 | 0.94 | 6 | 25.8 | 6 |
| Molise | 1.58 | 29 | 45.0 | 25 | 3.53 | 98 | 53.6 | 83 |
| Urbino 'Carlo Bo' | 1.58 | 27 | 33.3 | 4 | 1.14 | 15 | 31.3 | 11 |
| Napoli 'Federico II' | 1.51 | 25 | 45.6 | 29 | 2.03 | 75 | 45.7 | 46 |
| Teramo | 1.45 | 23 | 54.8 | 75 | 2.42 | 90 | 64.7 | 96 |
| Cattolica del Sacro Cuore | 1.45 | 21 | 50.0 | 52 | 1.29 | 25 | 45.8 | 48 |
| Salento | 1.45 | 19 | 66.7 | 98 | 1.03 | 11 | 25.0 | 4 |
| Foggia | 1.45 | 17 | 45.1 | 27 | 1.65 | 54 | 49.1 | 61 |
| Bari | 1.42 | 15 | 48.1 | 38 | 1.27 | 23 | 35.1 | 21 |
| 'Campus Bio-medico' | 1.36 | 13 | 53.8 | 69 | 1.93 | 67 | 55.0 | 86 |
| Rome 'Tor Vergata' | 1.34 | 11 | 37.1 | 11 | 1.34 | 29 | 34.4 | 15 |
| Turin | 1.33 | 8 | 47.1 | 36 | 1.43 | 38 | 50.0 | 67 |
| Insubria | 1.33 | 6 | 34.8 | 6 | 3.21 | 94 | 44.8 | 42 |
| Bologna | 1.27 | 4 | 42.7 | 19 | 1.78 | 63 | 52.7 | 75 |
| Cagliari | 1.21 | 2 | 52.9 | 63 | 2.12 | 81 | 68.4 | 98 |
| Trieste SISSA | 0.85 | 0 | 22.2 | 0 | 0.87 | 2 | 11.1 | 0 |
| Total | 2.72 | - | 49.5 | | 1.73 | | 45.4 | |

*Table 4: Effectiveness of overall mobility by Italian universities (sorted by M 1.1)*

|   | R 1.1 | R 1.2 | R 2.1 | R 2.2 | T 1.1 | T 1.2 | T 2.1 | T 2.2 | M 1.1 | M 1.2 | M 2.1 | M 2.2 | Productivity |
|---|---|---|---|---|---|---|---|---|---|---|---|---|---|
| R 1.1 | - | | | | | | | | | | | | |
| R 1.2 | 0.25 | - | | | | | | | | | | | |
| R 2.1 | 0.05 | 0.40 | - | | | | | | | | | | |
| R 2.2 | 0.09 | 0.55 | 0.73 | - | | | | | | | | | |
| T 1.1 | 0.06 | -0.01 | -0.09 | -0.13 | - | | | | | | | | |
| T 1.2 | 0.05 | 0.03 | -0.04 | -0.06 | 0.56 | - | | | | | | | |
| T 2.1 | -0.24 | -0.17 | -0.15 | -0.13 | 0.60 | 0.36 | - | | | | | | |
| T 2.2 | 0.02 | 0.03 | -0.02 | 0.00 | 0.32 | 0.63 | 0.43 | - | | | | | |
| M 1.1 | 0.92 | 0.17 | -0.04 | -0.03 | 0.33 | 0.19 | -0.06 | 0.09 | - | | | | |
| M 1.2 | 0.21 | 0.82 | 0.32 | 0.40 | 0.22 | 0.46 | -0.06 | 0.30 | 0.21 | - | | | |
| M 2.1 | -0.10 | 0.22 | 0.49 | 0.39 | 0.30 | 0.31 | 0.58 | 0.42 | -0.01 | 0.26 | - | | |
| M 2.2 | -0.06 | 0.29 | 0.47 | 0.70 | 0.05 | 0.27 | 0.12 | 0.49 | -0.07 | 0.36 | 0.59 | - | |
| Productivity | -0.25 | 0.02 | 0.47 | 0.45 | -0.02 | -0.10 | -0.11 | -0.14 | -0.26 | -0.02 | 0.06 | 0.20 | - |

*Table 5: Spearman correlation matrix*

# 5. Conclusions

Given the importance of human capital in the current knowledge-based economy, organizations compete to attract, develop and retain the best available talent in the labor market. Competitive institutions strive to optimize and improve the effectiveness of their processes of recruitment and career advancement. In the higher education system, such processes are fundamental not only for the competitiveness of the institutions themselves but also in terms of the role of the university in support of industrial competitiveness, socio-economic development and social mobility. In competitive higher education systems, world-class universities compete with one another to bring in the best researchers and teaching professors, from both home and abroad. In less competitive systems, as seen in many European nations, the incentive to do so is less noticeable. In those nations, including the Italian one, where scarce competition is associated to high levels of favoritism, merit is not always the prime criteria for faculty selection..In Italian universities which operate under rigid administrative structures, the dismissal of a professor is exceptionally rare, and turnover is only voluntary. In cases where governments have pursued improvement, progress is slow and meets notable resistance. One of the instruments to limit favoritism could be to monitor the effectiveness of universities' recruitment and turnover, and making it part of incentive systems. In Italy, both the second (VQR 2004-2010) and the third national research evaluation exercise (VQR, 2011-2014) in fact introduced such an incentive: a "bonus" for research products authored by professors recruited by the university over the period of observation, and achieving "excellent" scores. In spite of the methodological weaknesses of the VQR (Abramo and D'Angelo, 2015), and the still marginal character of the linked funding, the added attention to recruitment is an interesting element. The work presented here relates to this and other national contexts, proposing a multidimensional system for monitoring and evaluating the effectiveness of recruitment and turnover processes.

The proposed indicators offer two perspectives of evaluation: one is internal, on the scale of the university; the other is external, at national scale. The first perspective evaluates the recruitment/turnover relative to the university's potential capacity to attract and retain human resources, in terms of the research quality of the existing faculty. The second evaluates the effectiveness in absolute terms, by comparing the research performance of the recruited/separated professors to the entire population of Italian professors.

The empirical validation of the proposed indicators is carried out in application to the Italian academic system, analyzing and comparing the scientific productivity of all recruited and separated professors over the 2008-2012 period, for the fields where bibliometric evaluation techniques are considered reliable.

The analyses for the study context show that, for Italian universities, there is no significant correlation between the effectiveness of recruitment and effectiveness of turnover. However there are relevant correlations between the indicators of effectiveness of recruitment and indicators of overall effectiveness of mobility, and to a lesser degree between indicators of overall effectiveness and effectiveness of turnover. In addition, the analyses encounter a situation where, at the national scale, there is no correlation between the productivity of the universities and their capacity to retain their talented scholars. However, again at national scale, there is a relevant correlation between the universities' productivity and their effectiveness in recruitment.

The indicators presented can be used by the universities for *ex-post* monitoring and evaluation of their own processes of recruitment and turnover, as objectives measures of their effectiveness. The application of such measures could also assist in incentivizing merit-based processes of recruitment and turnover, reducing phenomena of favoritism and nepotism.

The research presented offers new elements with respect to the established literature, in the range and specificity of human resources practices analyzed, as well as for breadth of disciplines observed.